\documentclass[openacc=False]{rstransasub}

\usepackage{xspace}
\definecolor{orange}{rgb}{1, 0.5, 0}

\newcommand{\dcc}{P2400196-v6}

\newcommand{\gw}[1][]{GW#1\xspace}
\newcommand{\elmag}{electromagnetic (EM)\renewcommand{\elmag}{EM\xspace}\xspace}
\newcommand{\snr}[1][]{signal-to-noise ratio#1 (SNR#1)\renewcommand{\snr}[1][]{SNR##1\xspace}\xspace}

\newcommand{\pdet}{p_\mathrm{det}}
\newcommand{\pfd}{p_\mathrm{FD}}
\newcommand{\pfa}{p_\mathrm{FA}}
\newcommand{\hyp}{\mathcal{H}}
\newcommand{\unlens}{\mathrm{U}}
\newcommand{\lens}{\mathrm{L}}
\newcommand{\noise}{\mathrm{N}}
\newcommand{\HU}{\hyp_\unlens}
\newcommand{\HL}{\hyp_{\lens}}
\newcommand{\HN}{\hyp_{\noise}}

\titlehead{Review}

\begin{document}

\title{False positives for gravitational lensing: the gravitational-wave perspective}

\author{
David Keitel$^{1,2}$}

\address{$^{1}$Departament de F\'isica, Universitat de les Illes Balears, IAC3--IEEC, Crta. 
Valldemossa km 7.5, E-07122 Palma, Spain \\
$^{2}$University of Portsmouth, Institute of Cosmology and Gravitation,
Portsmouth PO1 3FX, United Kingdom
}

\subject{astrophysics, relativity}


\corres{David Keitel\\
\email{david.keitel@ligo.org}\\[\baselineskip]
18 July 2024
-- \href{https://dcc.ligo.org/\dcc}{LIGO-\dcc}}

\begin{abstract}
For the first detection of a novel astrophysical phenomenon,
scientific standards are particularly high.
Especially in a multi-messenger context,
there are also opportunity costs to follow-up observations on any detection claims.
So in searching for the still elusive lensed gravitational waves,
care needs to be taken in controlling false positives.
In particular, many methods for identifying strong lensing rely on
some form of parameter similarity or waveform consistency,
which under rapidly growing catalog sizes can expose them to false positives
from coincident but unlensed events
if proper care is not taken.
And searches for waveform deformations in all lensing regimes
are subject to degeneracies we need to mitigate between lensing,
intrinsic parameters,
insufficiently modelled effects such as orbital eccentricity,
or even deviations from general relativity.
Robust lensing studies also require understanding and mitigating
glitches and non-stationarities in the detector data.
This article reviews sources of possible false positives
(and their flip side: false negatives)
in gravitational-wave lensing searches
and the main approaches the community is pursuing to mitigate them.
\end{abstract}


\begin{fmtext}
\vspace{-1.5\baselineskip}
\section{Introduction}

Its sensitivity increasing with each observing run,
the LIGO--Virgo--KAGRA (LVK) network
of gravitational-wave (GW) detectors
\cite{LIGOScientific:2014pky,VIRGO:2014yos,KAGRA:2018plz}
is reaching deeper into the universe
and opening up chances for new types of detections.
One exciting opportunity are gravitationally lensed signals,
with a diverse science case as exposed in other papers in this issue,
e.g.~\cite{Smith:2024mml}.
At present, lensing effects are primarily of interest
for \gw[s] from compact binary coalescences.

\end{fmtext}


\maketitle

\begin{figure}[!h]
\centering\includegraphics[width=0.75\textwidth]{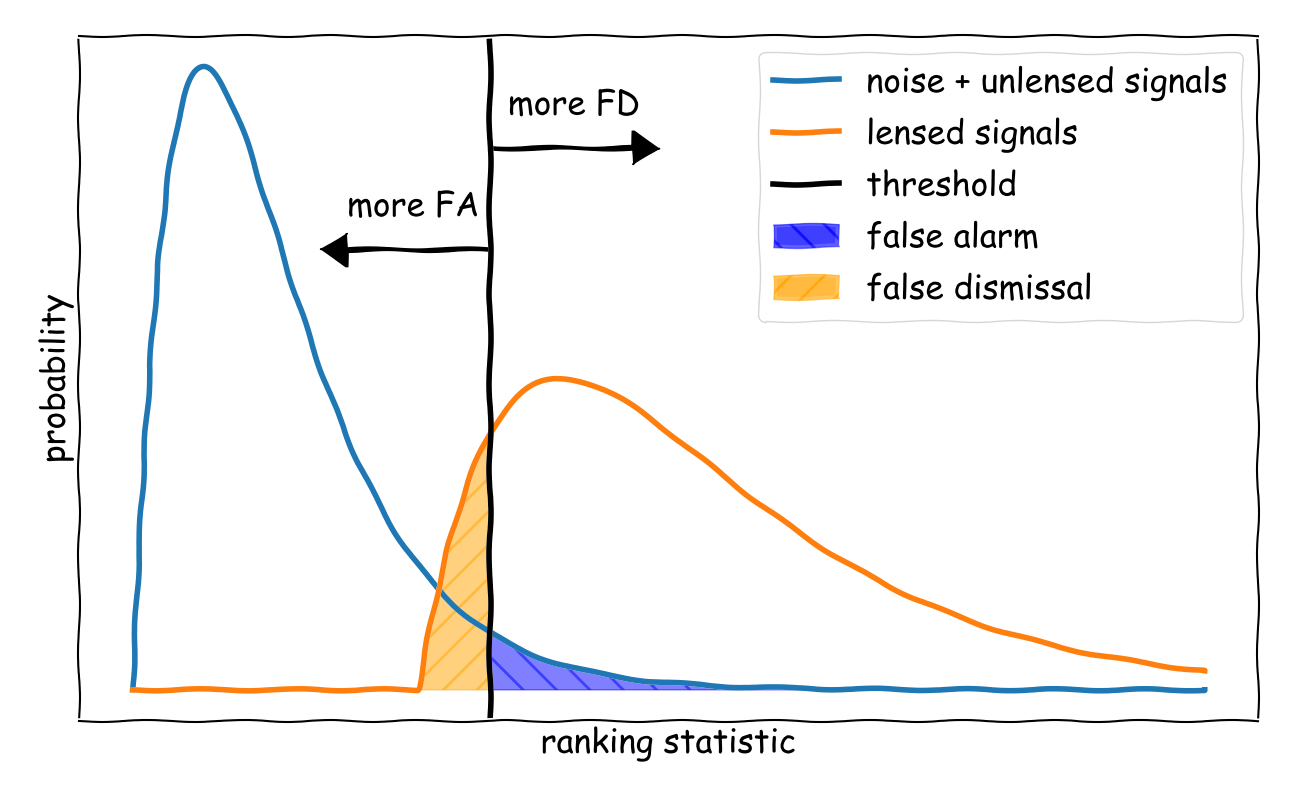}
\caption{Sketch of the basic hypothesis test dilemma:
lowering a detection threshold on some ranking statistic
increases false alarms (FA)
while raising it increases false dismissals (FD).
}
\label{fig:pfa_pdet}
\end{figure}

No lensed \gw[s] have been detected so far
\cite{Hannuksela:2019kle,LIGOScientific:2021izm,LIGOScientific:2023bwz},
and only a small fraction of events
is expected to have noticeable lensing signatures --
on the order of one in hundreds or thousands for the current detector generation
\cite{Ng:2017yiu,Li:2018prc,Oguri:2018muv,Wierda:2021upe,Xu:2021bfn}.
So our problem is
finding one of these rare lensed events
in the rapidly growing catalogues of all transient \gw detections
\cite{KAGRA:2021vkt}.
This corresponds to the well-known basic dilemma of all hypothesis tests,
as illustrated in \autoref{fig:pfa_pdet}:
we want to achieve a decent detection probability $\pdet$
(corresponding to a low false-dismissal or false-negative probability \mbox{$\pfd = 1-\pdet$})
at limited false-alarm or false-positive probability $\pfa$.

As a snapshot, as of writing of this article (2024-05-24)
already 98 significant new candidates from the ongoing O4 observing run
had been reported\footnote{\url{https://gracedb.ligo.org}},
in addition to the 90 signals reported from O1--O3~\cite{KAGRA:2021vkt},
With future sensitivity improvements~\cite{KAGRA:2013rdx},
the rate will grow further with accessible cosmic volume,
i.e. the third power of detector noise curve improvements.

Our key challenge for lensed \gw[s] is thus
finding needles in a haystack, while the haystack rapidly grows.
This has to be met without the help of
the most informative aspect of lensing studies
in \elmag astronomy: image geometry,
since \gw sky localisation is far worse~\cite{KAGRA:2013rdx}
than typical image sizes and separations.
Instead, we can rely on the LVK detectors,
with their Hz--kHz sensitive range,
to provide high time resolution
and to extract the source parameters encoded in the
amplitude and phase evolution of the
\gw strain time series (the ``waveforms'')
that we can typically track across many cycles.

This paper will discuss the various sources of $\pfa$
in identifying lensed \gw[s],
and the solutions found so far
or under active development in the \gw lensing community.
The discussion will be split into two cases:
(i) identification of sets of strongly-lensed multiple images
in the geometric regime;
(ii) deformed waveforms from
strongly lensed type II images
and the beating patterns or wave-optics effects
commonly referred to as ``milli''- and ``microlensing'' of \gw[s]
\cite{Deguchi:1986zz,Liu:2023ikc}.

\section{Multiple images}
\label{sec:pairs}

In the geometric optics regime,
strong lensing produces multiple images of the same source
with almost identical waveforms
(up to magnification, time delays,
and phasing changes that are only observable in the presence of strong higher modes
\cite{Ezquiaga:2020gdt}).
So these can be identified by checking pairs
(or higher multiples)
of \gw events for consistent
sky location
and intrinsic parameters.
Actually, this is a triple hypothesis test:
lensed images of a single source ($\HL$)
vs. unrelated unlensed signals ($\HU$)
vs. at least one of the candidates being just a noise fluctuation ($\HN$).
But most methods developed so far
(posterior overlap \cite{Haris:2018vmn},
machine-learning tools \cite{Goyal:2021hxv,Magare:2024wje},
phase consistency \cite{Ezquiaga:2023xfe}
and joint parameter estimation \cite{Liu:2020par,Lo:2021nae,Janquart:2021qov})
typically focus on the binary $\HL$ vs $\HU$ test,
assuming that the candidates have already been identified
as astrophysical over terrestrial noise
with sufficient confidence
by standard \gw searches~\cite{KAGRA:2021vkt}.
This section will review the main sources of $\pfa$
for identifying multiply imaged \gw[s],
and methods to mitigate them.

\subsection{False positives from pure noise masquerading as GWs?}

\begin{figure}[!h]
\centering\includegraphics[width=\textwidth]{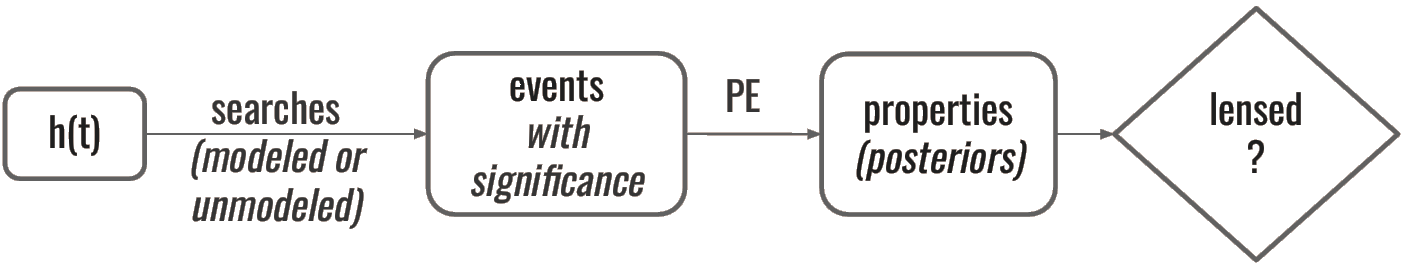}
\caption{Simplified flowchart of the detection process
from \gw strain data, $h(t)$,
via search pipelines and
parameter estimation (PE)
towards lensing hypothesis tests.
}
\label{fig:flowchart}
\end{figure}

In a simplified workflow of detecting \gw candidates and performing lensing hypothesis tests,
as illustrated in \autoref{fig:flowchart},
the most basic source of false positives is mistaking noise fluctuations
for astrophysical signals.
This can be considered as mostly under control
given the long development and sophistication of \gw search algorithms.
In particular, significances assigned by matched-filter pipelines
are not just based on \snr[s] under a Gaussian noise assumption.
Instead, ranking statistics incorporate a variety of information and vetoes
on signal shape, data-quality information, etc.
(e.g. \cite{Allen:2004gu,Dhurandhar:2017aan,Godwin:2020weu}).
The backgrounds used to assign significance are
empirically estimated from the actual data.
See appendix D of~\cite{KAGRA:2021vkt},
earlier GWTC papers\cite{gwtc1,gwtc2,gwtc21},
and references therein.

One caveat to consider is that the thresholds for inclusion in a GWTC
are not actually extremely strict --
GWTC-3 cut at \mbox{$p_\mathrm{astro} > 0.5$}
and has an estimated noise contamination of $\sim$10--15\%
\cite{KAGRA:2021vkt}.
But in fact,
several methods have been developed
for \emph{sub-threshold searches},
trying to dig out even fainter lensed images from below the GWTC thresholds
that match other, stronger events.
They reduce the covered parameter space,
and hence the statistical trials factor,
based on the target events
\cite{McIsaac:2019use,Li:2019osa,Dai:2020tpj,Li:2023zdl}.
To avoid false positives from pure noise,
extra care needs to be taken in studying candidates
from such searches
from the data quality perspective.
Additional confidence can also be gained from studying
the consistency of lensing parameters for candidate image sets
\cite{Goyal:2023lqf,Ng:2024ooy,Janquart:2023mvf}.

\subsection{False positives from unlensed GWs masquerading as lensed pairs?}

Likely the main issue for pair identification
are false alarms from coincidentally similar
but physically independent \gw events.
While the chance of actual lensed events only grows linearly,
the set of possible coincident pairs grows quadratically with catalog size,
even for uniformly distributed parameters.\footnote{
Higher multiples will be revisited later.}
In addition, it has become clear already
that merging binary black holes
cluster in some regions of mass and spin space
\cite{KAGRA:2021duu}.
(While for binaries including neutron stars,
the numbers are still too low for detailed population studies.)

But does the $\pfa$ actually grow as rapidly as the number of all event pairs?
The relevant quantity is the catalogue-level total:
$1-(1-\pfa^\mathrm{per\,pair})^{N_\mathrm{pairs}}$.
This problem has been studied quantitatively in \cite{Caliskan:2022wbh,Wierda:2021upe}.
A simplified approximation to real lensing pipelines
was used in \cite{Caliskan:2022wbh},
choosing an operating point of a per-pair $\pfa$ of $10^{-4}$
from checking overlap in mass, sky location and phase.
From this, they found the global $\pfa$
to rapidly approach unity beyond 100 detections,
i.e. from O4 on.
However, there are several ways out of this apparent dilemma.
\cite{Haris:2018vmn} had previously already demonstrated
that time-delay priors can significantly decrease $\pfa$
-- this will be discussed in detail in a moment.
Additionally,
\cite{Caliskan:2022wbh} pointed out that
increasing \snr thresholds
and analysing triple or quad images
(also see below)
can reduce the scaling of $\pfa$ with catalogue size.
As discussed further below,
both \elmag follow-up
(in rare cases)
or additional smoking-gun \gw-only signatures of lensing
would be very safe solutions.
And even without such extra sources of confirmation,
the actual \gw lensing analysis pipelines
can be much stricter in rejecting chance coincidences,
reducing the per-pair $\pfa$ --
especially full joint parameter estimation
\cite{Liu:2020par,Lo:2021nae,Janquart:2021qov}
is more restrictive than overlap statistics,
and more so when restricting to specific lens models~\cite{Janquart:2022zdd}.
See section \ref{sec:pairs-better} for further discussion of these solutions.

A crucial ingredient to reducing the steep scaling
of the catalogue-level $\pfa$ is to consider time delay priors.
As demonstrated already by \cite{Haris:2018vmn}
for the posterior overlap method,
multiplying its Bayes factor with another
that compares the priors for the relatively short time delays from galaxy lenses
against coincident unlensed pairs
can reduce the per-pair $\pfa$ from $10^{-2}$ to $10^{-5}$
at a fixed $\pdet\approx0.8$.
Furthermore, \cite{Wierda:2021upe}
have pointed out that,
at fixed per-pair $\pfa$,
the catalogue-level growth
(with increasing observing time
at constant detection rate)
is suppressed from quadratic to linear,
as in single-event searches.
As further pointed out by \cite{Wierda:2021upe},
for cluster lenses with their larger allowed time delays
the situation is more difficult,
but the scaling can still be suppressed as long as
the expected lensed time delays are typically shorter
than the span of the whole catalogue.

In summary,
the per-pair $\pfa$ can be reduced significantly by several considerations,
and the scaling of overall false positives with catalogue size
can also be controlled, at least for galaxy lenses,
by considering astrophysical time-delay priors.

\subsection{False positives from extra noise making GWs look lensed?}

Combining the two previous concerns,
it might happen that one or more real \gw events
are contaminated by additional noise transients
to such a degree that they falsely resemble others and hence appear lensed.
The impact in particular of large glitches
on \gw parameter estimation is known,
and deglitching methods are crucial \cite{Davis:2022ird}.
See for example GW170817~\cite{LIGOScientific:2017vwq}
(where clean subtraction was possible)
and GW200129\_065458
(where subtraction uncertainties are significant~\cite{Payne:2022spz,Macas:2023wiw}).
For pair identification,
the situation is no different than in other areas of \gw analysis
and should not be a dominant additional source of $\pfa$.
Section~\ref{sec:singles}\ref{sec:singles-noise} will revisit this
for frequency-dependent single-event lensing signatures.

\subsection{False positives from waveform systematics?}

Understanding \gw[s] from compact binaries relies on
waveform modelling~\cite{Schmidt:2020ekt}.
The existing model families demonstrate strong agreement
in the ``vanilla'' parameter-space regions
(similar masses, low spins, quasi-circular),
but open challenges remain for
large mass ratios,
high precessing spins,
and orbital eccentricity.
Imperfect waveforms can lead to
shifted posteriors or missed modes,
and hence to both false alarms and false dismissals
in lensing pair identification.

\gw lensing studies so far have generally used
the IMRPhenomXPHM model~\cite{Pratten:2020ceb}
or earlier versions from the same family.
A first dedicated study~\cite{Garron:2023gvd}
of the impact of waveform choice on lensing searches
has covered the posterior overlap method~\cite{Haris:2018vmn}.
After identifying that some problematic cases
turned out to be due to sampler convergence issues in the GWTC releases
rather than actual waveform systematics,
no issues were identified on O1--O3 events.

For O4 and beyond with increased sensitivity,
the higher \snr[s] for some events
and increased chances of detecting non-vanilla signals
mean waveform systematics can become more important.
On the other hand,
continued waveform development
will allow for better-constrained posteriors
and hence reduce the lensing $\pfa$
from coincident pairs.

\subsection{Improvements with higher multiples}

Moving on to solutions for reducing $\pfa$,
as already discussed,
the $\pfa$ from coincident unlensed pairs drops steeply
when more than two images of the same source can be identified
\cite{Caliskan:2022wbh}.
Triples and quads also allow for detailed checks
of the time delays, phase differences, and their ordering
against the known classification of possible configurations of lens geometries
\cite{Wierda:2021upe,Ng:2024ooy},
especially when combined with full joint parameter estimation
and explicit lens model choices~\cite{Janquart:2022zdd}.

However, even if there are three or four reasonably bright images,
we may not detect all of them.
First, the \gw detectors are simply not in observing state
for non-negligible times,
with duty factors of e.g. 53\% for both LIGO detectors together
or 83\% for at least one of them online during O4a\footnote{
\url{https://gwosc.org/detector_status/O4a/}
-- for O4b Virgo has joined again and improved this}.
Second, even slightly lower relative magnifications
or changes in the detector noise level
can push some images below the typical detection thresholds.
Sub-threshold searches
\cite{McIsaac:2019use,Li:2019osa,Dai:2020tpj,Li:2023zdl}
can thus be particularly valuable,
in particular when a promising candidate pair or triple is already known
and can be used to narrow down the list for possible thirds and fourths further
\cite{Dai:2020tpj,Ng:2024ooy}.

\subsection{Improvements with smoking guns}

Another solution to any $\pfa$--$\pfd$ balancing problem
is to find additional, clear signatures of the effect searched for
-- smoking guns, or in this case maybe ``smoking magnifying glasses''.

One case for \gw lensing are mergers involving neutron stars.
Even without an \elmag counterpart,
from \gw[s] alone,
the matter effects on the waveform
(tidal deformability changes to the phasing)
can break the degeneracy between
nearby heavy sources
and lensed far-away lighter sources
\cite{Pang:2020qow},
because lensed cases
would seem to contain heavy objects
with the stronger tidal deformability of a lighter one.
Tidal information is mainly encoded at higher frequencies
than the optimum of current detectors,
but ongoing sensitivity improvements
will make such checks feasible
for increasing numbers of low-mass candidates.

For binary black holes,
smoking guns are still possible to find,
as will be discussed in \autoref{sec:singles}.
Strong lensing $\pfa$ can thus be reduced
when we also find evidence on one or more of the images of
the parity change in type-II images
\cite{Ezquiaga:2020gdt,Janquart:2021nus,Wang:2021kzt}
or of combined strong plus microlensing
\cite{Shan:2023ngi,Mishra:2023ddt}.

\subsection{Improvements with combined GW+EM observations}

In addition to the various \gw-only methods of decreasing $\pfa$,
a lensed \elmag counterpart could be considered a beyond-a-reasonable-doubt confirmation
of candidate lensed \gw[s],
but identifying such counterparts has its own challenges.
This appears attractive for binary neutron stars like GW170817~\cite{LIGOScientific:2017ync}
where a counterpart can be expected from the exact same source~\cite{Smith:2022vbp},
but the depth of \elmag observations can be a limiting factor
as well as the actual distinction of multiple images.
Specific cases like lensed neutron star -- black hole mixed binaries \cite{Magare:2023hgs}
and where the counterpart is a fast radio burst \cite{Singh:2023hbd}
have also been discussed in the literature.

However, given the relative detection reaches and redshift-dependent merger rates
\cite{Buscicchio:2020bdq,Buscicchio:2020cij},
lensing candidates involving neutron stars and with possible direct \elmag counterparts
will remain much rarer than those from binary black holes.
For those,
the ``counterpart'' is only a host galaxy,
with no temporally well-constrained \elmag signal
to make up for the broad \gw sky maps.
Strongly-lensed signals are thus special
in making such associations at least potentially feasible \cite{Hannuksela:2020xor,Wempe:2022zlk}:
The improved sky localisation from joint analysis of multiple \gw images
could narrow down the list of possible hosts to a few.
Then the right one can be identified
through consistency of the properties of the possible lenses
and of the lens profile reconstructed from the \gw image set.
If the \gw lensing properties hint at a cluster lens,
deep \elmag follow-up observations~\cite{Smith:2018kbc,Bianconi:2022etr}
can also probe the small number of such lens candidates
even within much broader single-event sky maps.

\subsection{Improvements with better data and better analyses to the rescue}
\label{sec:pairs-better}

To mitigate the various sources of $\pfa$ discussed above
even in the absence of \gw-only or \gw-\elmag smoking guns,
we can also consider the benefits of more sensitive or cleaner detector data
and of further improvements to our analysis methods.

On the detector side,
the improved sensitivity of O4 and runs beyond it
yields events with higher \snr[s],
which in turn yield narrower posteriors,
and hence at least a subset of strong candidates
can be identified with higher significance and lower $\pfa$
-- with the caveat of waveform systematics.
To maintain clean catalogues and parameter estimation,
it is also crucial for
glitch identification and mitigation
\cite{Davis:2022ird}
to keep up with instrumental improvements.

Meanwhile, methods development
for identifying lensed pairs (and higher multiples)
is still ongoing.
The initial candidate identification
via posterior overlap~\cite{Haris:2018vmn}
can be complemented with independent pipelines
using machine learning~\cite{Goyal:2021hxv,Magare:2024wje}
or phase information~\cite{Ezquiaga:2023xfe},
and candidates from sub-threshold searches
can be submitted to various follow-up steps
\cite{Wong:2021lxf,Goyal:2023lqf,Ng:2024ooy,Janquart:2023mvf}.
The final criterion is then generally taken to be
full Bayesian joint parameter estimation
\cite{Liu:2020par,Lo:2021nae,Janquart:2021qov},
which as discussed above can be significantly stricter than overlap-type analyses,
and can be further improved through lens model considerations~\cite{Janquart:2022zdd}.
More quantitative work is needed
to study false alarms under this full meta-pipeline of \gw lensing searches,
but it is clear that it can significantly improve the picture.

Efforts are also ongoing to better include expectations
from lens simulations and \elmag surveys
on the time delay, magnification and phase distributions
into initial candidate selection
\cite{Wierda:2021upe,More:2021kpb,Ng:2024ooy}.
One open issue with these is
that typically simple galaxy lenses are assumed,
with the distributions for clusters more difficult to model.
Hence, any low-$\pfa$ methods tailored to galaxy lenses
may incur significant $\pfd$ for other lenses.
Still, such approaches can be crucial tools
for a high-purity sample of \gw lensing candidates.

\section{Deformed waveforms (single images)}
\label{sec:singles}

Leaving behind the case of multiple images from strong lensing,
the main ways for identifying single lensed \gw images
make use of frequency-dependent deformations:
(i) the parity change in strong lensing type-II images
(detectable in the presence of higher modes and precessing spins
\cite{Ezquiaga:2020gdt,Janquart:2021nus,Wang:2021kzt});
and (ii) wave-optics effects~\cite{Deguchi:1986zz}
or the beating patterns between overlapping short-time-delay images
\cite{Liu:2023ikc}
both happening for low-mass lenses
and hence commonly referred to as ``micro''- and ``millilensing'',
despite their differences from the namesake \elmag phenomena.

These effects are generally searched for with Bayesian evidence calculation
(e.g. \cite{Janquart:2021nus,Wright:2021cbn,Liu:2023ikc,Mishra:2023ddt}).
In this case,
our haystack only grows linearly with the number of catalogue events.
As discussed before,
these signatures can also serve as additional smoking guns
to confirm strongly lensed multiple images.
However, they are themselves subtle to detect on signals
with realistic \snr[s]
and could be potentially confused with non-lensing effects,
if proper care is not taken.
This section will now focus on the specific sources of $\pfa$
for detecting such deformations themselves.
For most of these, investigations have only started recently,
so the overview will be quite brief and not yet quantitative;
full studies with updated waveform models and analysis techniques
will likely also discover ways of breaking these degeneracies
at least partially.

\subsection{False positives from noise issues?}
\label{sec:singles-noise}

As discussed for pair identification,
deglitching~\cite{Davis:2022ird} is crucial
for robust parameter estimation on any \gw events,
and only more so if we want to detect effects
beyond standard waveforms.
Mistaking noise effects for lensing
is an example of the general ``out-of-manifold'' effect
for matched-filtering based hypothesis tests:
with the likelihood based on a Gaussian noise assumption,
anything unusual in the data is likely to make
a nested test of different signal hypotheses
prefer the more complex hypotheses,
which have greater freedom to fit the extra noise.

Beyond mitigation by case-by-case noise characterisation and cleaning,
extensive background studies
with simulated signals in a variety of noisy data
are crucial
for setting robust thresholds on lensing Bayes factors.
This can reduce $\pfa$,
but at the obvious cost of a higher $\pfd$.
On the other hand,
ongoing developments for statistically stronger
and more model-informed analysis methods
are also likely to increase the noise-signal separation.

\subsection{False positives from waveform systematics?}

The same ``out-of-manifold'' issue,
but on the signal hypothesis side,
manifests as imperfect underlying waveform models
for unlensed \gw signals
increasing the chance for elevated lensing Bayes factors.
This can be due to a simple lack of precision
in the numerical-relativity calibration of the models,
but becomes more prominent when entire physical effects
are present in the real data but missing in the model.

In particular,
standard models have evolved from aligned-spin dominant-mode only models
to ``PHM'' versions
(including precessing spins and higher modes)
\cite{Varma:2019csw,Pratten:2020ceb,Ramos-Buades:2023ehm}
but are still limited in their implementations of precession.
and cover only quasi-circular binaries,
with modelling of orbital eccentricity still in an early phase
(e.g. \cite{Islam:2021mha,Ramos-Buades:2021adz,Liu:2023dgl,Liu:2023ldr}).
Studies of waveform systematics
i   n single-image lensing tests have started
\cite{Janquart:2023mvf},
but more systematic work is still needed.

\subsection{False positives from degeneracies with spin precession?}

Even if we had perfect waveform models
e.g. including spin precession and microlensing together,
there is a level of intrinsic degeneracy between the two effects.
If we consider \gw microlensing ``for dummies'',
the effect comes down to extra modulations (``wiggles'')
in the time series of \gw strain.
Phenomenologically, precession can cause similar modulations.
This was first studied on a set of example cases
by~\cite{Kim:2023scq},
finding that the two effects
indeed can look similar under realistic conditions
but can be distinguished with full parameter estimation.
A more systematic exploration of the full relevant parameter space
has been started in~\cite{Mishra:2023ddt}
and is being further extended in ongoing work,
which should allow to narrow down the regions of parameter space
where degeneracies are significant.

\subsection{False positives from degeneracies with orbital eccentricity?}

Similarly to spin precession,
orbital eccentricity can also cause extra waveform ``wiggles''
that can be mistaken for other effects.
As a famous example, not lensing-related,
the very short signal GW190521~\cite{LIGOScientific:2020iuh}
can be fit with a variety of modifications beyond
a standard quasi-circular model,
including with eccentricity (e.g. \cite{Romero-Shaw:2020thy}).
While several models now exist that incorporate eccentricity to some degree
(e.g. \cite{Islam:2021mha,Ramos-Buades:2021adz,Liu:2023dgl,Liu:2023ldr}),
much more work is needed for better calibration
and for treating full spin effects at the same time.
A first study on the possible degeneracy of eccentricity
with microlensing \cite{Mishra:2023ecc}
has found the risk
of unlensed signals from eccentric binaries
being mistaken for lensed signals from quasi-circular sources
to be potentially significant,
though eccentricities at the required levels
are likely rare in the overall source population.

\subsection{False positives from (or for) violations of general relativity?}

Yet another possible source of ``extra wiggles''
could be deviations from general relativity,
with ways to test for such violations
in \gw signals reviewed in \cite{Colleoni:2024lpj}.
However, this subsection can remain short,
as it can be argued that lensing is less exotic than a new theory of gravity
and hence that we should be more worried about
lensing as a source of false alarms for those~\cite{Gupta:2024gun}
rather than about such violations as false alarms for lensing.
For completeness,
first studies in that direction
already have already been performed
for both type-II images and microlensing
\cite{Ezquiaga:2022nak,Mishra:2023vzo,Wright:2024mco}.

\section{Conclusions}

In classical astronomy,
there are nowadays many photos
from which it is completely obvious that lensing is happening.
By contrast,
the reason to carefully consider risks of false positives in searches for \gw lensing
stems from the limited \snr[s] achievable with current detectors
and the lack of geometric information,
which is the richest observable in the \elmag sky.
For \gw lensing,
care needs to be taken to extract the maximum information
imprinted on the \gw strain time series.
The \gw lensing community
is thus focusing on developing robust methods
for obtaining high-confidence results.
To achieve these,
we must study and control all
imaginable sources of false alarms
and pursue concrete mitigation strategies.

As discussed in this brief review,
searches for multiple strongly lensed images
may primarily be susceptible to false alarms
from the much larger number of unlensed signals
forming coincidentally similar pairs.
Additional possible issues include
noise fluctuations
and waveform systematics.
For frequency-dependent deformations on single images,
possible sources of false positives again include
noise fluctuations
and waveform systematics,
as well as degeneracies with intrinsic compact binary parameters
such as spin precession and orbital eccentricity.

Many of these challenges
can be at least partially addressed by improvements
to detector sensitivity,
mitigation of noise issues,
waveform modelling,
and, in particular, dedicated \gw lensing analysis techniques.
Multi-image searches
can profit enormously from
using full joint Bayesian parameter estimation,
identifying more than two images of the same source,
incorporating specific lens models
to constrain time delays and image set configurations.
Through these avenues,
the scaling of false alarms
with the number of detections can be significantly reduced
\cite{Haris:2018vmn,Wierda:2021upe,More:2021kpb}
at least for galaxy lenses with their shorter time delays.
For some candidates,
the combination with additional smoking guns
such as matter effects in binaries including neutron stars
or the single-image signatures
of type-II images or microlensing
can yield additional evidence.
For the latter,
improved methods may be able to partially break degeneracies,
and systematic studies can narrow down the relevant parameter space regions
and in turn increase confidence in candidates not falling into these.

Multi-messenger lensing studies
can provide further inputs to reduce \gw lensing false alarms,
both through population-level studies
that allow for more restrictive \gw analysis techniques
and through counterpart searches for individual sources.

Another important conclusion is
the importance of background studies.
While \gw lensing hypothesis tests
are typically formulated in a Bayesian way,
we can still treat the resulting Bayes factors
as frequentist detection statistics
and calculate their background distribution
through simulation studies,
as done for the main LVC/LVK lensing studies so far
\cite{LIGOScientific:2021izm,LIGOScientific:2023bwz}.
To do this optimally,
one needs to sample over all possible confounding factors --
full population models,
spin precession,
orbital eccentricity,
and the diverse features of real detector noise.
For a proper statistical sampling,
such studies easily become more expensive
than the main analyses of actual detected \gw events,
but are a crucial ingredient for robust \gw lensing candidate identification.

It is also important to consider the expected evolution
of observational \gw astronomy
in the future.
As of the writing of this article,
the ongoing O4 run has produced over a hundred new significant \gw candidates,
and might double that by its end.
The O5 run is expected to produce daily detections~\cite{KAGRA:2013rdx}.
This increased rate also corresponds to a deeper reach into the universe;
thus, the chance for lensed signals to be in the data increases.
To profit from this increase,
the growth of false positives
must be controlled through the approaches outlined above.
KAGRA~\cite{KAGRA:2018plz}
and LIGO India~\cite{Unnikrishnan:2023uou}
extending the global network
will improve sky localisation~\cite{KAGRA:2013rdx},
reducing the risk of chance coincidences
and improving the chances for multi-messenger studies.
Many of the considerations reviewed here
also apply to next-generation detectors
with their truly cosmological reach,
such as the ground-based
Einstein Telescope~\cite{Punturo:2010zz}
and Cosmic Explorer~\cite{Evans:2021gyd},
or LISA in space~\cite{LISA:2017pwj},
though their regime will be completely different:
lensing will still be a rare phenomenon relatively,
but cease to be rare in absolute terms.
Additionally,
high \snr[s]
and long observation times thanks to lower observable frequencies
should help
lensing identification.

\ack{
Thanks to the members of the LIGO--Virgo--KAGRA gravitational lensing group
as well as the organisers and attendants
of the Royal Society meeting on Multi-messenger Gravitational Lensing
(Manchester, 11-12 March 2024)
for many fruitful discussions.
This work was supported by the Universitat de les Illes Balears (UIB);
the Spanish Agencia Estatal de Investigaci{\'o}n grants
CNS2022-135440,
PID2022-138626NB-I00,
RED2022-134204-E,
RED2022-134411-T,
funded by MICIU/AEI/10.13039/501100011033,
the European Union NextGenerationEU/PRTR,
and the ERDF/EU;
and the Comunitat Aut{\`o}noma de les Illes Balears
through the Direcci{\'o} General de Recerca, Innovaci{\'o} I Transformaci{\'o} Digital
with funds from the Tourist Stay Tax Law
(PDR2020/11 - ITS2017-006)
as well as through the Conselleria d'Economia, Hisenda i Innovaci{\'o}
with grant numbers
SINCO2022/6719
(European Union NextGenerationEU/PRTR-C17.I1)
and SINCO2022/18146
(co-financed by the European Union
and FEDER Operational Program 2021-2027 of the Balearic Islands).
The Royal Society
and the Institute of Cosmology and Gravitation (University of Portsmouth)
have provided additional travel support.
This material is based upon work supported by NSF's LIGO Laboratory
which is a major facility fully funded by the National Science Foundation.
\autoref{fig:pfa_pdet} was generated
using matplotlib~\cite{Hunter:2007mpl},
numpy~\cite{Harris2020:npy},
scipy~\cite{Virtanen:2020sci},
and \url{https://github.com/ipython/xkcd-font/}.
This paper has been assigned the document number \href{https://dcc.ligo.org/\dcc}{LIGO-\dcc}.
}

\bibliographystyle{vancouver}
\bibliography{lensFP}

\end{document}